\documentclass[twocolumn, preprintnumbers, showpacs, aps, nofootinbib]{revtex4}
\usepackage{graphicx}

\newcommand{\bi}{\bibitem}
\newcommand{\be}{\begin{eqnarray}}
\newcommand{\ee}{\end{eqnarray}}

\begin{document}

\title{A note on the black hole information paradox in de Sitter spacetimes}

\author{Cosimo Bambi}

\affiliation{Department of Physics and Astronomy, 
Wayne State University, Detroit, MI 48201, USA\\
Michigan Center for Theoretical Physics, 
University of Michigan, Ann Arbor, MI 48109, USA}

\date{\today}

\preprint{WSU-HEP-0804}

\begin{abstract}
The possibility of stable or quasi--stable Planck mass black 
hole remnants as solution to the black hole information paradox 
is commonly believed phenomenologically unacceptable. Since we 
need a black hole remnant for every possible initial state, 
the number of remnants is expected to be infinite and that would 
lead to remnant pair production in any physical process with a 
total available energy roughly exceeding the Planck mass. In 
this note I point out that a positive cosmological constant of 
the Universe would naturally lead to an upper bound on the number 
of possible remnants.
\end{abstract}


\maketitle

\section{Introduction}

The Black Hole (BH) information paradox is a well known issue 
in theoretical physics~\cite{reviews, volume}. In quantum mechanics, 
the evolution operator $U$ is unitary, i.e. $U^\dag U = 1$, 
which implies that a pure state must evolve into another pure 
state. On the other hand, BHs have a temperature and emit radiation
via Hawking effect~\cite{hawking}. This radiation is almost thermal 
and thus does not depend on the features of the initial state, but 
only on the BH geometry outside the horizon (BH mass, electric 
charge and angular momentum). In other words, the outgoing 
radiation does not carry any information about the initial state 
which collapsed into the BH. As the BH emits radiation, it loses 
mass. The crucial question is what happens at the end: if the BH 
evaporates completely and the final product is only thermal 
radiation, information is lost forever. In this case, a pure state 
would evolve into a mixed one, i.e. its evolution would be non--unitary. 
There are basically three answers to the question on the fate of 
the initial information: $i)$ the evolution is indeed non--unitary, 
$ii)$ the radiation is not thermal and carries information, $iii)$ 
information is stored in a long--lived or stable BH remnant. All 
the proposals, however, have some problems.

The possibility that {\it information is lost} and unitarity
is violated was originally put forward in ref.~\cite{prop-1}. 
In this case, the basic ingredients of quantum mechanics must 
be revised, in order to allow for a pure state to evolve into 
a mixed one. In particular, the unitary S--matrix of quantum 
field theory, which connects any initial quantum state to its 
final quantum state, $|\psi_{out} \rangle = S \, |\psi_{in}\rangle$, 
is replaced by the non--unitary super--scattering operator \$, 
which maps an initial density matrix into a final density matrix
\be
\rho_{out} = \$ \rho_{in}
\ee 
and mixes quantum states. The problem is that energy conservation
is violated and that there is no empty vacuum as ground state, 
which means we should live in a Universe similar to a thermal 
bath at the Planck temperature~\cite{crit-1}.

The second possibility is that {\it information is encoded in the 
outgoing radiation}~\cite{prop-2} and unitarity is conserved. 
Such an idea relies on possible higher order effects which are not 
taken into account in the standard calculations. Here we need new 
physics which is radically different from the one we know and 
``sacred'' concepts such as the ones of locality and causality must 
be abandoned. At present, this is likely the most appealing 
possibility, even because supported by the AdS/CFT conjecture. 
Nevertheless, one could expect that departure from standard 
calculations becomes important only when the BH approaches the 
Planck mass: for large BHs, the radiation would be well described 
by the semi--classical framework, while, for small BHs, new
physics would play and important role and the emitted radiation 
could carry all the information about the initial state. However, 
general arguments suggest that it is unlikely a sudden emission of 
the information in the last stages of evaporation~\cite{prop-3}. The 
BH should emit about $N \sim M^2 / M_{Pl}^2$ quanta with average 
wavelength $\lambda \sim N/M_{Pl}$, where $M$ is the initial BH 
mass and $M_{Pl}$ is the Planck one. On the other hand, the size 
of the BH at this stage is about $1/M_{Pl}$, which means that the 
wave--function overlap is $1/N^3$ and the simultaneous emission 
of $N$ quanta would be suppressed by the factor $1/N^{3N}$. Even 
in the case of a gradual emission of $N$ quanta, the final process 
would require a very long time, at the level of $N^4/M_{Pl}$. So, 
the end product of Hawking evaporation would be a long--lived 
remnant, which is the last option.

{\it Information is stored into a stable or quasi--stable 
remnant}~\cite{prop-3}. This possibility is suggested by the idea 
that semi--classical calculations of BH evaporation  hold till the 
BH approaches the Planck mass, when quantum gravity and back reaction 
effects can no longer be neglected. Here the problem is that an
infinite number of initial states implies an infinite number of 
remnants. Even assuming a tiny coupling constant between remnants 
and ordinary matter, virtual remnant states should affect all the 
quantum processes. Remnant pair production should be the result of 
any reaction with a total available energy exceeding the remnant 
mass~\cite{crit-3}: for macroscopic processes, this is easy to 
achieve and, even if the remnant production probability is extremely 
small, mainly because the energy densities are always small, the 
existence of infinite states would be anyway catastrophic.

\section{Upper bound on the number of remnants}

The argument against BH remnants is that an infinite number of them
would be phenomenologically unacceptable~\cite{crit-3}. Every 
initial state must be associated with a remnant and if we think 
that the number of initial states is infinite, the number of 
remnants must be infinite too.

The aim of this note is to show that a positive cosmological
constant $\Lambda$ can provide a natural upper bound on the number 
of remnants. Cosmological and astrophysical evidences suggest 
that in the Universe $\Lambda > 0$~\cite{data}. If this were the 
case, the Universe would be asymptotically de Sitter and we could 
estimate the corresponding entropy. For an an empty de Sitter 
spacetime with a cosmological constant equal to the one we can 
deduce from observations, the entropy is~\footnote{In the case of 
non--empty de Sitter spacetime, the entropy is smaller, so 
eq.~(\ref{entropy-ds}) can be seen as an upper bound.}
\be\label{entropy-ds}
S_0 = \frac{3 \, \pi \, M_{Pl}^2}{\Lambda} 
\sim 2 \cdot 10^{122}\, .
\ee
In ref.~\cite{banks} was conjectured that $S_0$ is roughly the 
logarithm of the total number of quantum states necessary to describe 
the Universe and therefore the direct count of the number of its 
degrees of freedom. Following this interpretation, the total number 
$\mathcal{N}$ of possible initial states which can collapse into a 
BH is bounded by
\be\label{number}
\mathcal{N} < \mathcal{N}_{bound} = 
\exp \left( \frac{3 \, \pi \, M_{Pl}^2}{\Lambda} \right)
\sim 10^{10^{122}} \, .
\ee
Let us note that this is only an upper bound and that actually the
number of remnants may be even much smaller.

\begin{figure}[t]
\par
\begin{center}
\includegraphics[width=8cm,angle=0]{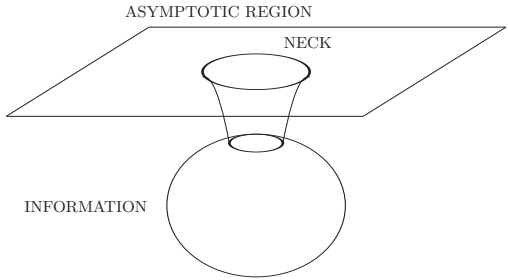}
\end{center}
\par
\vspace{-7mm} 
\caption{Information could be stored within a large
volume that appears of Planck size for an external 
observer living in the asymptotic region.}
\label{fig}
\end{figure}

One could also be worried about the fact that a 
huge amount of information is contained in an object with a Planck 
volume. This is not a fundamental problem, because the spacetime is 
curved and a small volume for an external observer could be a large or 
infinite volume inside the BH~\cite{volume}. An intuitive picture of 
how this is possible is sketched in fig.~\ref{fig}.

\section{Dangerous processes}

Even if there is not yet any reliable theory in which one can discuss
quantum gravity objects like BH remnants, we can anyway try to 
estimate some implications of remnants from ``reasonable'' arguments.

For example, it is common belief that we can take quantum gravity 
effects into account in low--energy particle physics by using an 
effective field theory, with non--renormalizable and Planck mass 
suppressed operators. Here virtual BH remnants could induce baryon 
and lepton number violations and, in particular, proton 
decay~\cite{p-decay}. In 3+1 dimensions, proton decay would be 
described by the dimension six operator
\be\label{op-6}
\mathcal{O}_6 \sim \frac{\psi\psi\psi\psi}{M_{Pl}^2} \, ,
\ee
since two quarks of the proton are converted into a quark (or an 
anti--quark) and a charged lepton. The usual estimate of the proton 
lifetime is
\be
\tau_p \sim \frac{M_{Pl}^4}{m_p^5} 
\sim 10^{45} \; {\rm yr}
\ee
and has to be compared with present experimental limits, where 
several channels are bounded by $\tau_p > 10^{33}$~yr~\cite{pdg}.
If remnants exist, every remnant intermediate state is equally
probable and integrating over an infinite number of states the
decay width goes to infinity. In our case their number is instead
finite and the proton lifetime is
\be
\tau_p \gtrsim \frac{1}{g^2 \, \mathcal{N}_{bound}} \, 
\frac{M_{Pl}^4}{m_p^5} \, ,
\ee
where $g$ is the dimensionless coupling constant between ordinary 
matter and BH remnants which we put in front of the dimension six 
operator in eq.~(\ref{op-6}). If the number of remnants is not much
smaller than the upper bound in eq.~(\ref{number}), the coupling 
constant $g$ has to be really small. However, the possibility of
a tiny coupling constant cannot be rejected, because this is just
an effective theory. There could be indeed mechanisms capable of
suppressing the process. A proposal to get a small coupling can be 
found, for example, in ref.~\cite{nir}.

As for the remnant pair production in macroscopic physical
processes, the estimate is even more difficult, albeit the
phenomenon is based on more solid arguments. Here we have 
to expect some suppression factor due to the ratio between 
the remnant external volume and the macroscopic volume 
which provides the necessary amount of energy to create the 
remnant pair. Anyway, the problem still reduces to find a 
small coupling constant between BH remnants and ordinary matter.

Very dangerous processes may be represented by gravitational 
phenomena, because the validity of the Equivalence Principle 
requires that all the form of energy couples to gravity with 
the same strength. For example, let us consider the $10^6$ Solar 
mass BH which is believed to reside at the center of the Galaxy. 
Like any BH, it is expected to emit Hawking radiation. For a 
Schwarzschild BH of mass $M$, the temperature $T$ is  
\be\label{temperature}
T = \frac{1}{8 \pi G_N M} = 5 \cdot 10^{-18} \, 
\left(\frac{10^6 \; M_\odot}{M}\right) \; {\rm eV}
\ee
and is so low that it seems to be impossible to observe, even in
the future. If remnants behave as ordinary particles, their emission 
would be Boltzmann suppressed by a huge Planck mass to BH temperature 
ratio, $\exp\left( - \frac{M_{Pl}}{T} \right) \sim 10^{-10^{45}}$.
If $\mathcal{N} \sim \mathcal{N}_{bound}$, they
should anyway cause a sudden evaporation of the BH at the center of 
the Galaxy. However, such argument is weak, because the
formation of remnants at the horizon may be more strongly suppressed:
remnants may not be ``elementary'' and, as we do not expect that the
emission of hydrogen atoms is similar to the one of protons, we
can argue that the same holds for remnants.

\section{Conclusions}

The black hole information paradox is an open issue which
clearly shows the difficulties to unify quantum mechanics 
and gravity in a unique framework. Several proposal have 
been suggested, but all of them present serious problems.

If we believe that semi--classical calculations are reliable 
as far as the mass of the black hole is much larger than the
Planck one, the most natural solution to the puzzle is that
information is stored in stable or long--lived remnants.
It is usually believed that there must exist an infinite 
number of remnants, one for every possible initial state, and 
this turns out to be phenomenologically unacceptable~\cite{crit-3}. 
In this note I showed that the number of remnants could instead
be finite and I provided an upper bound. The presence of a
positive cosmological constant would constrain the number of
degrees of freedom of the Universe and such a constraint can 
be seen at least as an upper bound on the number of initial 
states which can collapse into a black hole. That would make
the number of remnants finite. So, the main argument against 
remnants may not exist. If the number of remnants is close to 
the upper bound, it is still difficult to assert if they are 
phenomenologically acceptable. For sure, they cannot be 
completely excluded as solution to the information black hole 
paradox, because we do not how to describe them and, in the
framework of a low--energy effective field theory, they may
be associated with very small coupling constants.

\begin{acknowledgments}
I wish to thank Katherine Freese for kind hospitality 
at the Michigan Center for Theoretical Physics, 
where this work was being written.
The work is supported in part by NSF under grant PHY-0547794 
and by DOE under contract DE-FG02-96ER41005.
\end{acknowledgments}


\end{document}